# Computational analysis of short-range interactions between an edge dislocation and an array of equally-spaced identical shearable or non-shearable precipitates

Amirreza Keyhani

*The George W. Woodruff School of Mechanical Engineering, Georgia Institute of Technology, Atlanta, GA 30332-0405, USA*

**Abstract**

The interaction between dislocations and precipitates plays an important role in the mechanical behavior of alloys. To provide more insight into the physics of this interaction, this research analyzes short-range interactions of an edge dislocation with an array of equally-spaced identical precipitates. We use a modified dislocation dynamics approach accounting for penetrable and impenetrable precipitates. This research quantifies the effects of precipitate resistance on the geometry of the dislocation-precipitation interaction and the local distribution of plastic strain near a precipitate. The results show that a precipitate with a higher resistance causes an increase in the maximum value of dislocation curvature during the bypass. In addition, a higher level of precipitate resistance leads to a lower level of plastic deformation. Moreover, we observed a high plastic strain gradient at the interface of non-shearable precipitates.

Keywords: Dislocation Dynamics (DD), Penetrable and Impenetrable Precipitates, Dislocation-Precipitate interaction

## 1. Introduction

The development of computational methods for analyzing movements and interactions of dislocations has paved the way for the statistics of dislocations-precipitate interactions. Early simulation methods were geometrical and modeled precipitates as dimensionless obstacles against dislocation movements. The need for more realistic models motivated the development of dislocation dynamics (DD) [1-4]. This approach has been widely used for analyzing the physics of plasticity at micron scales. However, the simulation of dislocation-precipitate interaction in DD has been a challenging topic. Many devoted studies to dislocation-precipitate interactions are limited to only the stress fields caused by precipitates. While some studies [5-8] introduced precipitates as spherical stress fields, other studies [9-11] evaluated the stress field resulting from matrix and precipitate shear modulus mismatch by applying the superposition principle, which decomposes the dislocation-precipitate interaction into two problems: a dislocation problem in an infinite homogeneous body and a correction problem representing the elastic stress field of precipitates. Hence, the latter requires an extra-numerical method such as FEM or BEM. The coupling of DD and an extra computational method complicates the solution of large systems with a random distribution of precipitates.

The complications and disadvantages of modeling precipitates with the stress fields motivated Keyhani et al. [12-16] to develop a more efficient and robust methodology for modeling precipitates within the dislocation dynamics approach. This methodology uses a resistance scale to model precipitates. The present research applies this computational method to provide more insight into dislocation-precipitate interactions. This research quantifies the interaction between an edge dislocation and array of equally-spaced identical precipitates with various sizes and resistance levels. In addition, the author combines the modified dislocation dynamics approach and the finite element method to study the local distribution of plastic strain close to precipitates.

## 2. Modeling Approach

In this research, we use a recently proposed computational method [15] to model precipitates in the three-dimensional dislocation dynamics (DD) simulation code, DDLab [17]. Here, we briefly review the dislocation dynamics approach and the used methodology for modeling precipitates. The dislocation dynamics approach discretizes a dislocation curve into straight lines and defines each segment by its two end nodes. The mobility function $\mathbf{M}$ relates the vector of nodal forces $\{\mathbf{f}_i\}$ to the nodal velocity $\mathbf{v}_i$:

$$\mathbf{v}_i = \mathbf{M}\left(\{\mathbf{f}_i\}\right). \quad (1)$$

The velocity of node $i$, $\mathbf{v}_i$, depends on the forces acting on the other nodes. The mobility function depends on the orientations of the dislocation segments and material properties. We calculate the nodal velocities by solving mobility equations. Then, we compute the dislocation motion by topological considerations. More details on DD and other computational approaches that have so far been developed for determining the structure and motion of a single dislocation can be found in Ref. [17].

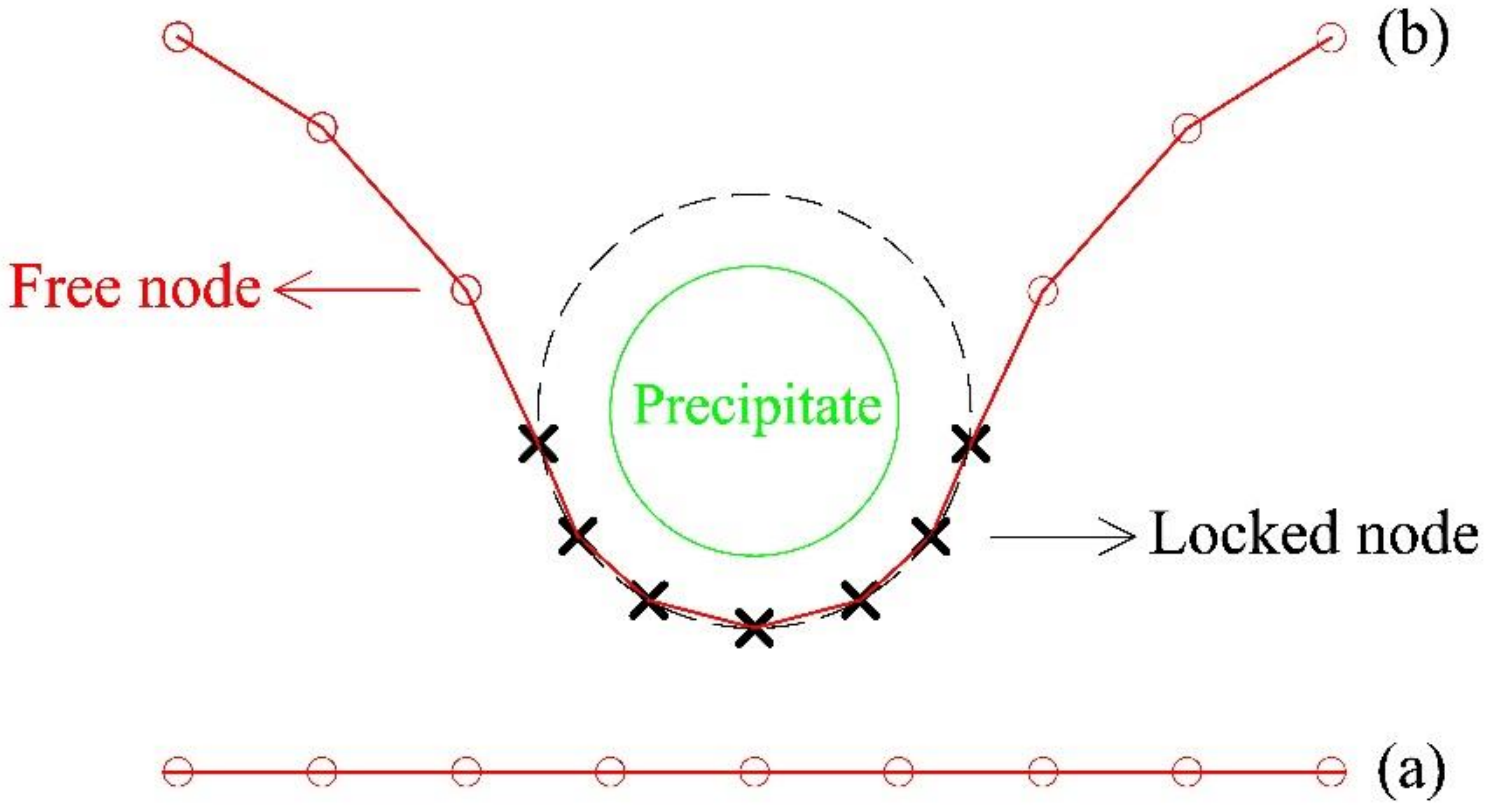

Fig. 1. (a) A dislocation line, (b) nodes that are positioned closer than a specific distance to the precipitate and that we locked. Circles and crosses represent the free and locked nodes, respectively.

To model precipitates in three-dimensional dislocation dynamics, we lock a dislocation node that is closer than a specific distance to a precipitate, as shown in Fig. 1. By this approach, since the dislocation segment pinned between two precipitates behaves similarly to the Frank-read source, we transform the main problem of the dislocation-precipitate interaction into the Frank-read source mechanism. To obtain equivalent results, the critical stress of two mechanisms must be equal. We assume that a dislocation rounds a precipitate with a larger modeling diameter $\left(D_m\right)$ than the precipitate diameter $\left(D\right)$. With equivalent Frank-read nucleation stress and modified Orowan stress $\tau_{\text{Orowan}} = \mu b \ln(D_1/r_0)/(2\pi L)$, the modeling diameter $\left(D_m\right)$ of a precipitate can now be determined (see Fig. 2):

$$L_f = L + D - D_m, \tag{2}$$

$$D_m = L + D - 2\pi L\beta\left[\ln\left(D_1/r_0\right)\right]^{-1}, \tag{3}$$

where $L$ is the internal distance between the two precipitates, $r_0$ is the core radius of dislocation, and $D_1 = (D^{-1} + L^{-1})^{-1}$.

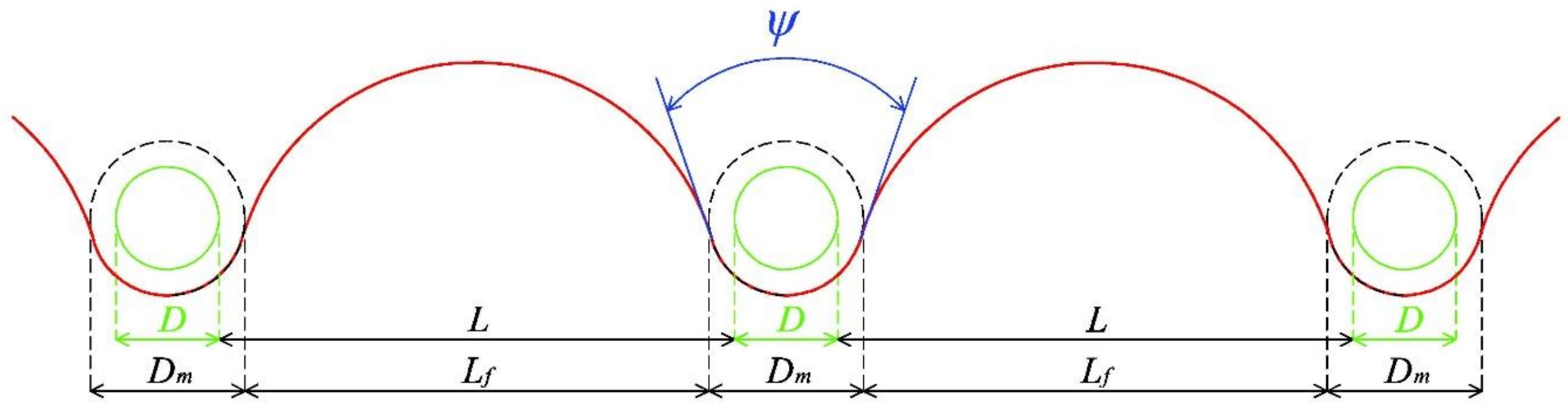

Fig. 2. A dislocation segment between two precipitates with internal distance $L$ acting as a Frank-Read source with length $L_f$. $\psi$ is the angle between the dislocation tangents on both sides of the precipitate.

The precipitate resistance $(R_{\mathrm{p}})$ is the level of shear stress required to cut the precipitate. When a dislocation encounters a precipitate, it bends, so the related shear stress that the dislocation exerts on a precipitate increases. If this stress reaches a critical value, the dislocation overcomes precipitates via two mechanisms: passing either by or through the precipitate. The first is when a dislocation forms a loop and passes by the highly resistant (impenetrable) precipitate. The maximum stress that a dislocation can exert on a precipitate is the point at which the dislocation radius of the curvature equals the modeling radius, so

$$\tau_{\max} = 2\mu b / D_m \tag{4}$$

whenever the precipitate resistance is higher or equal to this magnitude (i.e., $R_{\mathrm{p}} \geq \tau_{\max}$) and the dislocation stops behind the precipitate completely. This type of precipitate is called an "impenetrable precipitate." The second mechanism is when a dislocation passes through a precipitate, when the precipitate resistance is lower than the maximum stress (i.e., $R_{\mathrm{p}} \leq \tau_{\max}$). In this case, the dislocation passes through the precipitate by exerting a lower level of stress on the precipitate, which generates a radius of curvature larger than the modeling radius. This kind of precipitate is called a "penetrable precipitate." The applied stress on a dislocation must be sufficiently large to bend the dislocation to a critical position in order to pass by or through a precipitate.

The precipitate resistance scale $(R)$ is set to 1 for an impenetrable precipitate and 0 when no precipitate exists, with a linear interpolation between them:

$$R = \begin{cases} R_{\mathrm{p}} / \tau_{\max} & R_{\mathrm{p}} \leq \tau_{\max} \\ 1 & R_{\mathrm{p}} > \tau_{\max} \end{cases} . \tag{5}$$

When the distance of a node from a precipitate is less than the modeling radius, we lock the node, and at each step, compare the precipitate resistance to the local shear stress, the latter of which is related to the local curvature of this point. If the related local shear stress exceeds the precipitate resistance, we release the node.

## 3. Results and Discussion

This study analyzes several aspects of short-range interactions between which is initially an edge dislocation and an array of equally-spaced identical precipitates including (1) the evolution of the dislocation geometry while the dislocation passes by or through an array of precipitates at various resistance levels, (2) the effect of the precipitate resistance level on the overall effective plastic strain, and (3) the local distribution of effective plastic in a small domain around the precipitate.

### *3.1. Geometrical features of dislocation-precipitate interaction*

The geometry of a dislocation interacting with an array of precipitates is investigated in the Fe crystal (BCC) with a Burgers' vector $\mathbf{b} = [0.143 \quad 0.143 \quad 0.143]\,\mathrm{nm}$ in the $[\bar{1} \quad 0 \quad 1]$ glide plane. The mechanical properties are the shear modulus, $G = 81\,\mathrm{GPa}$, and Poisson's ratio, $\upsilon = 0.29$. The diameter of precipitates is 100 nm. We study the interaction over the range of 0.8 to 1.2 times the critical resolved shear stress (CRSS). Figures 3(a) and 3(b) illustrate the dislocation curvature and angle for each applied stress, respectively. Figure 3(a) shows that when the applied stress is lower than the critical state, the dislocation does not pass the precipitate completely, and the curvature and the angle remain constant after a specific time. When the applied stress becomes higher than CRSS, the dislocation curvature and the bypass angle reach their maximum and minimum values, respectively. After passing the precipitate, the dislocation curvature decreases while the bypass angle increases. Figures 4(a) and 4(b) illustrate the dislocation curvature and the bypass angle for three precipitate resistance ratios of 0.6, 0.8, and 1. The applied stress is equal to the critical stress for the impenetrable precipitate. Figure 4(a) shows that the maximum dislocation curvature decreases and the minimum dislocation angle increases after the precipitate resistance decreases.

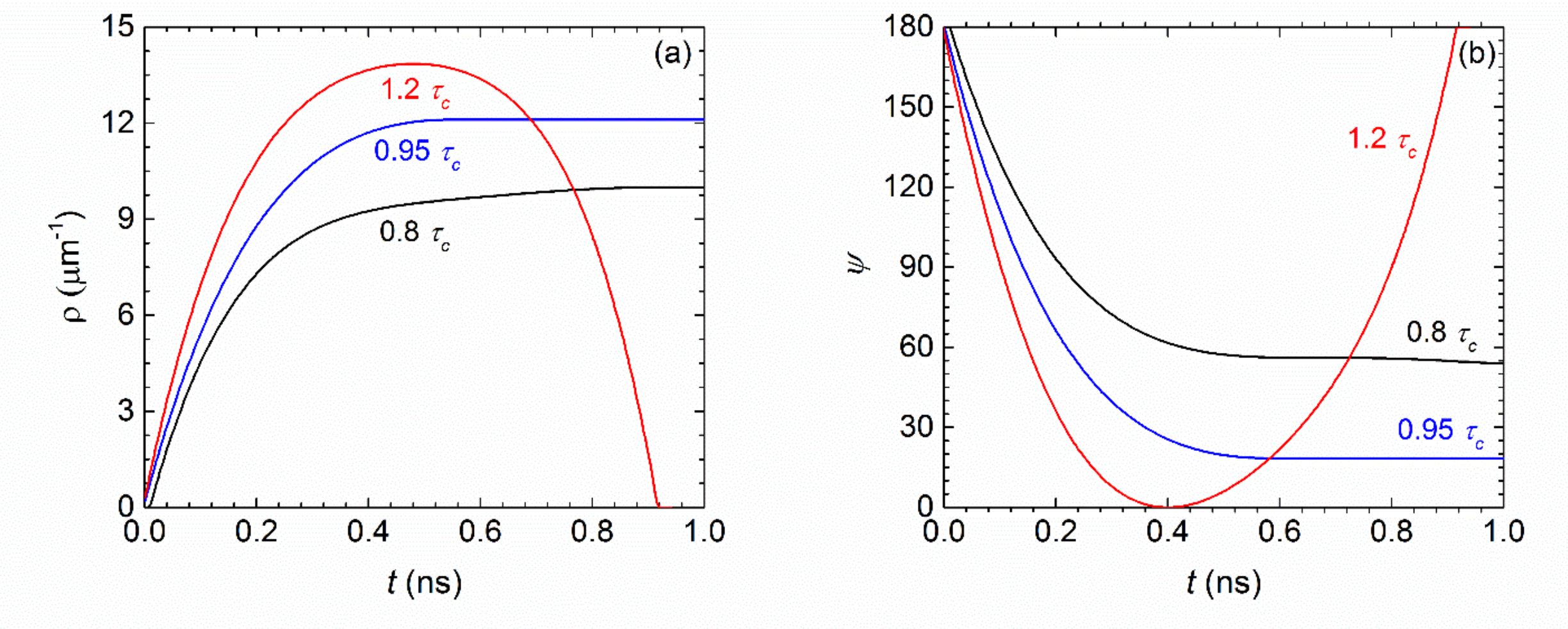

Fig. 3. (a) Variation in the dislocation curvature during interaction with an impenetrable precipitate, and (b) variation in the dislocation angle during interaction with an impenetrable precipitate.

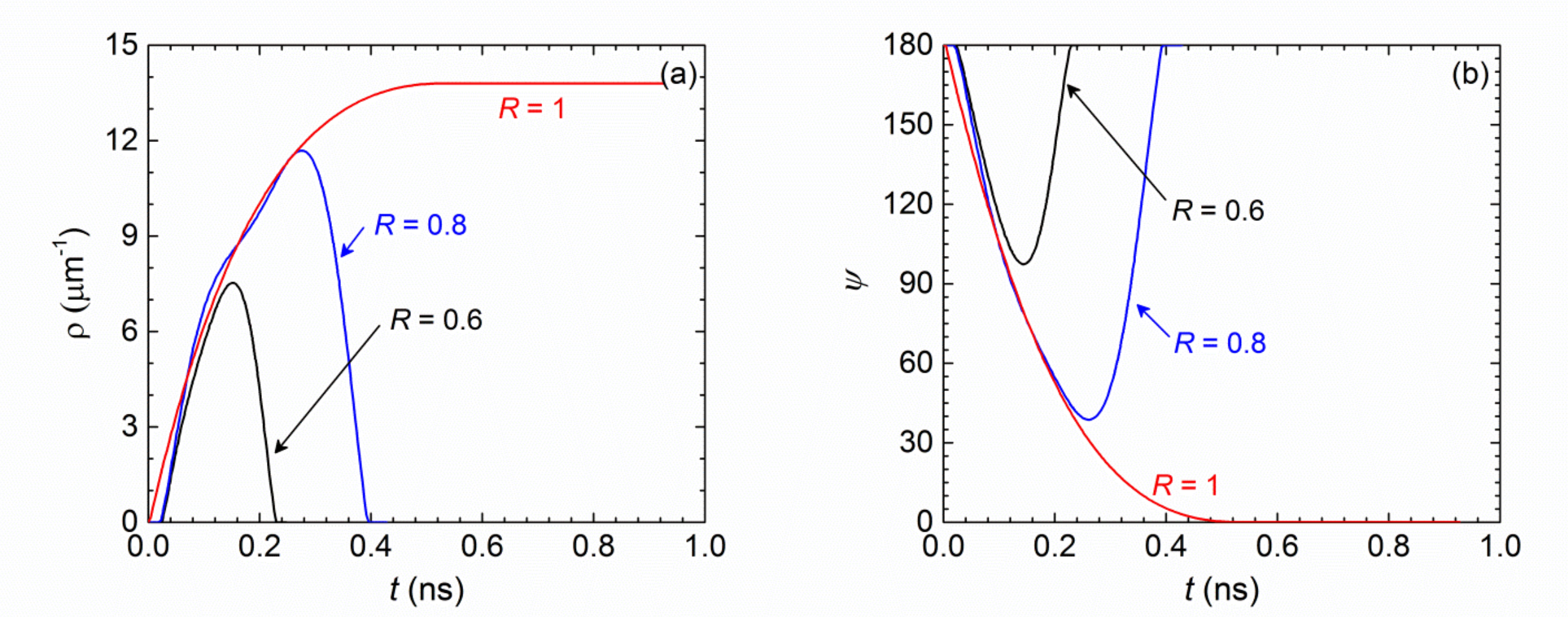

Fig. 4. (a) Variation in the dislocation curvature during interaction with a penetrable precipitate, and (b) variation in the dislocation angle during interaction with a penetrable precipitate.

### *3.2. Effect of precipitate resistance on effective plastic strain*

In this section, we analyze effective plastic strain over the process domain during the dislocation-precipitate interaction (Fig. 5) for various precipitate resistance ratios over the applied shear stress range of $0.8-1.2\,\tau_c$. The critical stress $\left(\tau_c\right)$ is the magnitude of the applied external stress when the dislocation is about to pass a non-shearable precipitate (critical state) but cannot pass it completely.

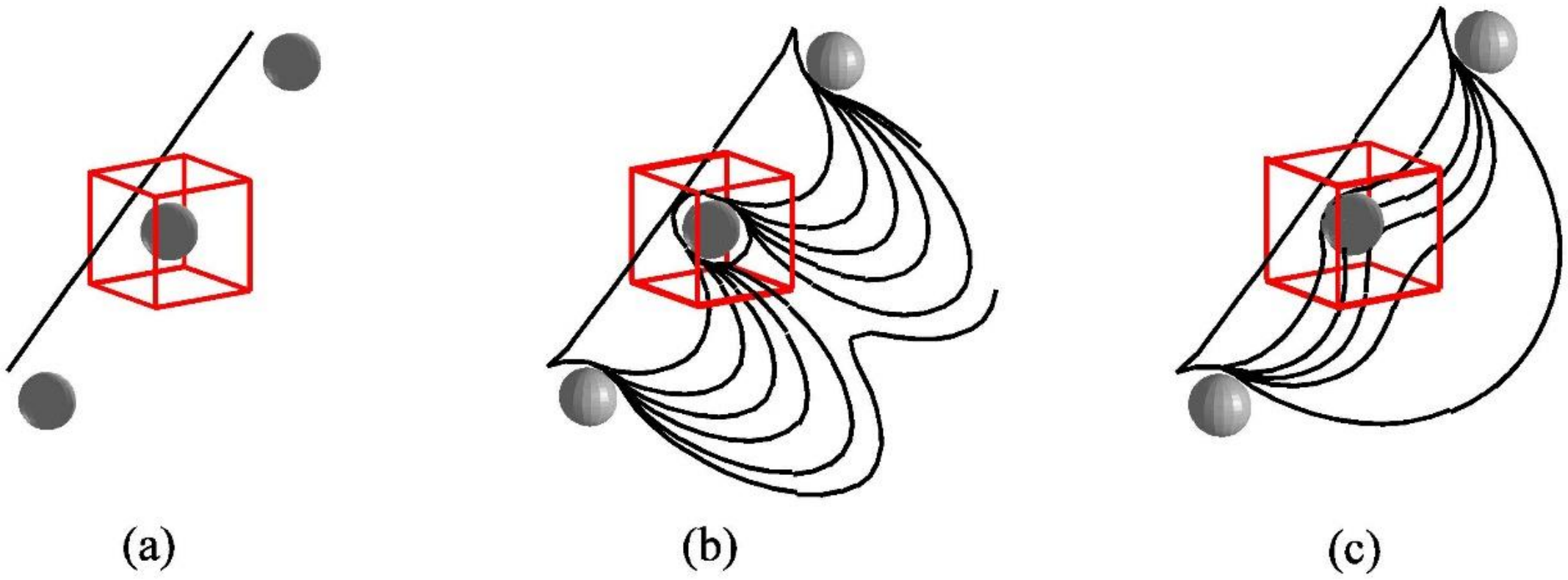

Fig. 5. Dislocation-precipitate interaction, (a) the process domain near a precipitate for computing the effective plastic strain, (b) dislocation encounters an impenetrable precipitate and loops around and passes by it (the Orowan mechanism), and (c) dislocation encounters a penetrable precipitate and passes through it by shearing.

Figure 6 illustrates the results of 40 simulations. This figure shows that the effective plastic strain decreases significantly when a precipitate resistance increases to a point where it completely stops its dislocation motion. In addition, it shows that either the applied stress on a dislocation is sufficiently large to pass a precipitate ( $\tau > \tau_c$ ) or not ( $\tau < \tau_c$ ), the effective plastic strain decreases once a dislocation encounters impenetrable precipitates (with a resistance ratio of 1). More importantly, it indicates that the induced effective plastic strain becomes independent of both the precipitate resistance ratio and the applied stress when a dislocation passes through by shearing (i.e., the precipitate resistance ratio is lower than 1). In the present simulation, the effective plastic strain is nearly $7\times10^{-4}$ when a dislocation passes a penetrable precipitate or when no precipitate exists. Obviously, the induced effective plastic strain nearby a penetrable precipitate is equal to an impenetrable precipitate when the applied stress is not sufficiently large to overcome the precipitate.

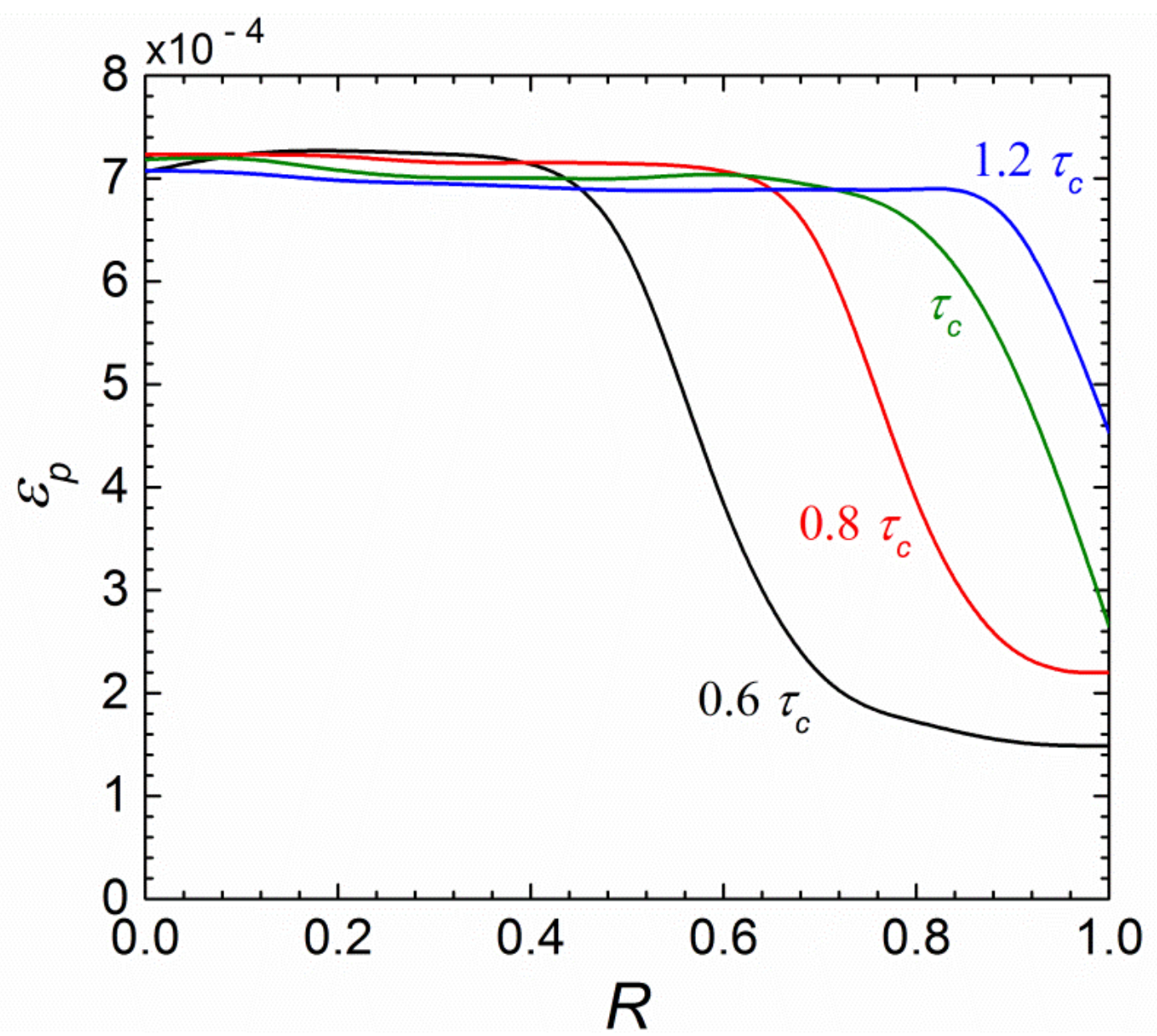

Fig. 6. The effective plastic strain versus the resistance ratio for various external stresses.

The difference between the effective plastic strains corresponding to $\tau = \tau_c$ and $\tau = 1.2\ \tau_c$ is related to the definition of critical stress. While a dislocation does not pass a precipitate completely when the applied stress is equal to the critical stress $(\tau = \tau_c)$, it passes an impenetrable precipitate even if the applied stress is a little larger than the critical stress $(\tau_c)$, so the resulting effective plastic strain increases because of the dislocation motion.

*3.3. Multi-scale analysis of plasticity close to a precipitate*

To show the plastic strain variation in a small domain near a precipitate, we use the multi-scale framework, originally proposed in Refs. [18, 19]. Adopting the finite element method on a macro scale, we calculate the plastic strain by means of explicit three-dimensional discrete dislocation dynamics on a micro-scale (Fig. 7). This hybrid approach allows us to address complex phenomena in deformation on a small scale. Neglecting inertia and damping terms, we conduct the coupling of two scales by the following equation:

$$\mathbf{KU} = \mathbf{f}_{\mathbf{ext}} + \mathbf{f}_{\mathbf{B}} + \mathbf{f}_{\infty} + \mathbf{f}_{\mathbf{P}}, \tag{6}$$

where $\mathbf{K}$ is the stiffness matrix and $\mathbf{U}$ is the nodal displacement vector, $\mathbf{f}_{\mathbf{ext}}$ is the applied force vector, and the other terms on the right-hand side are the force vectors related to the line dislocation dynamics method. Force vector $\mathbf{f}_{\mathbf{B}}$ arises from the long-range dislocation stress field and $\mathbf{f}_{\infty}$ is applied to treat boundary conditions as the dislocation stress field is usually in an infinite domain.

$\mathbf{f_p}$ arises from dislocation motions and results in an equivalent plastic strain in the finite element analysis,

$$
\begin{aligned}
\mathbf{f_{ext}} &= \int_{\Gamma} \bar{\mathbf{t}}\mathbf{N}\,d\Gamma \\
\mathbf{f_B} &= \int_{\Omega} \mathbf{S_D}\mathbf{B}\,d\Omega \\
\mathbf{f}_{\infty} &= -\int_{\Gamma} \mathbf{t}^{\infty}\mathbf{N}\,d\Gamma \\
\mathbf{f_p} &= \int_{\Omega} \mathbf{D}\boldsymbol{\varepsilon}_{\mathbf{p}}\mathbf{B}\,d\Omega
\end{aligned}
\qquad (7)
$$

in which $\bar{\mathbf{t}}$ and $\mathbf{t}^{\infty}$ are the applied traction and the resulting traction from the presence of dislocations on the boundary $\Gamma$, respectively. $\mathbf{S_D}$ is the average stress field resulting from the presence of dislocations in the finite domain, $\Omega$, which is identical to each element of the finite element analysis. $\mathbf{N}$ is the vector of shape functions, $\mathbf{B}\ \nabla{=}\mathbf{N}$ . $\boldsymbol{\varepsilon}^{\mathbf{p}}$ is the plastic strain vector resulting from dislocation motions, and $\mathbf{D}$ is the elastic stiffness tensor.

We modeled the dislocation-precipitate interaction in an infinite body and the effect of the long-range interaction by the direct dislocation-dislocation interaction algorithm of the standard DDLAB code. The same domain in Section 3.2 is discretized by $12\times12\times12$ ordinary cubic finite elements, and the bottom surface of the sample is fixed. As mentioned earlier, the generated mesh is not required to be consistent with the precipitate geometry since the developed methodology is independent of the continuum modeling (Fig. 7). The relevant effective plastic strain for the impenetrable and penetrable precipitates with resistance ratios of 0.6 are presented in Fig. 8(a) and Fig. 8(b), respectively. In order to show the effective plastic strains nearby precipitates more clearly, Fig. 9(a) and Fig. 9(b) illustrate the variations of effective plastic strain on the dislocation glide plane for impenetrable and penetrable precipitates, respectively.

The effective plastic strain presented in Fig. 9(b), which is related to the penetrable precipitate with a resistance ratio of 0.6, is not much higher than the one for the impenetrable precipitate in Figure 11(a). This is, however, in contrast to Fig. 6 in the first analysis which showed that the effective plastic strain reduces significantly by increasing the precipitate resistance. The application of the multi-scale modeling shows that although the effective plastic strain over the process domain containing a penetrable precipitate is higher than the domain which contains an impenetrable precipitate, the local effective plastic strain does not follow the same rule. In fact, Fig. 6 illustrates the average effective plastic strain in terms of the applied stress and the precipitate resistance over the whole process domain, whereas Fig. 9 shows the local distribution of effective plastic strain in the same domain. Although the change in the average effective plastic strain over the process domain is the result of the local effective plastic strain variation, it is impossible to predict the local variation just on the basis of the average over the process domain. Consequently, the multi-scale procedures are necessary to analyze the local effective plastic strain.

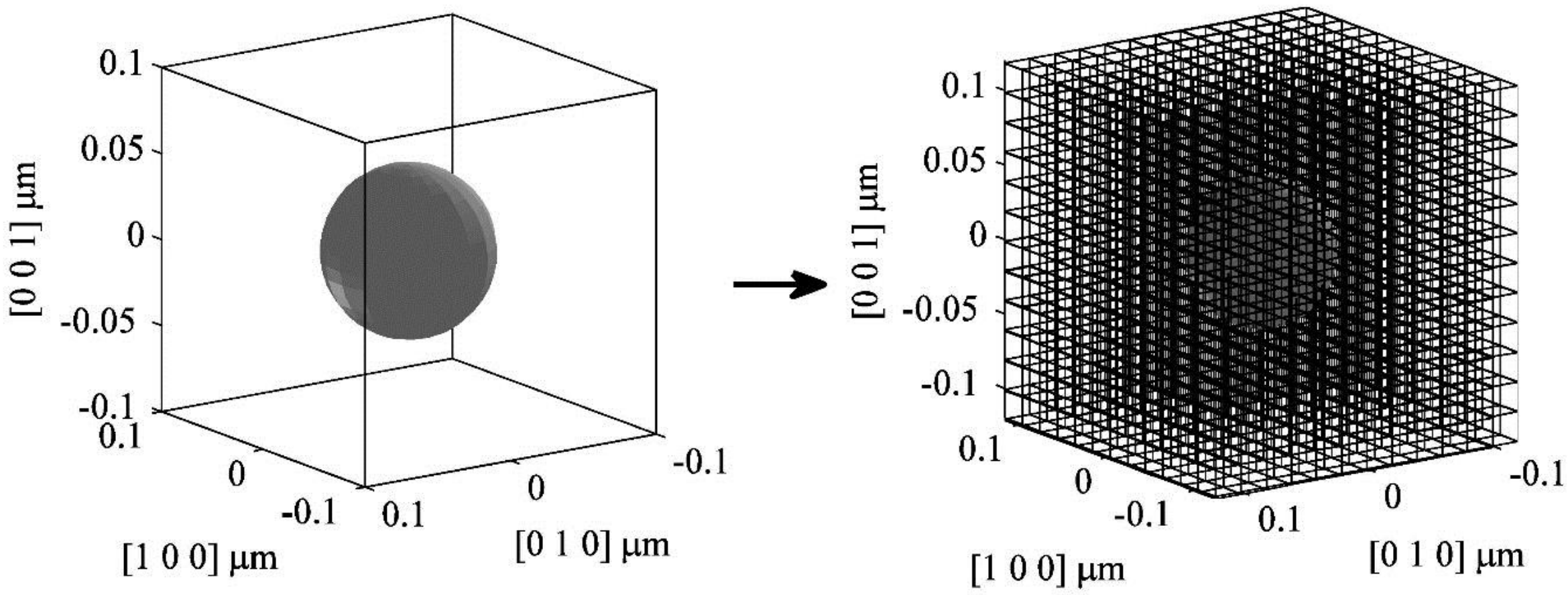

Fig. 7. The process domain is meshed with 12×12×12 ordinary 8-node cubic elements (independent from the precipitate geometry).

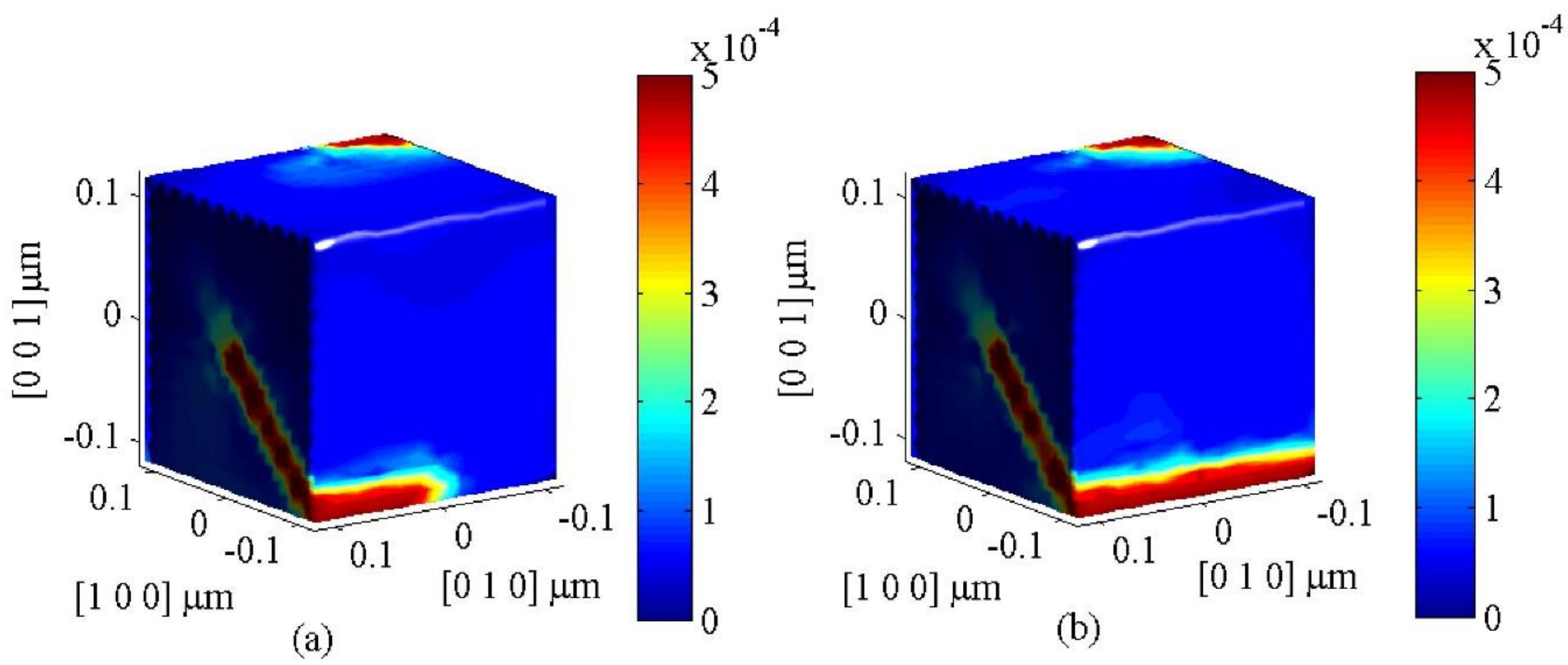

Fig. 8. Local effective plastic strain in the process domain, (a) impenetrable precipitate with a diameter of 100 nm, (b) penetrable precipitate with a diameter of 100 nm and a resistance ratio of 0.6.

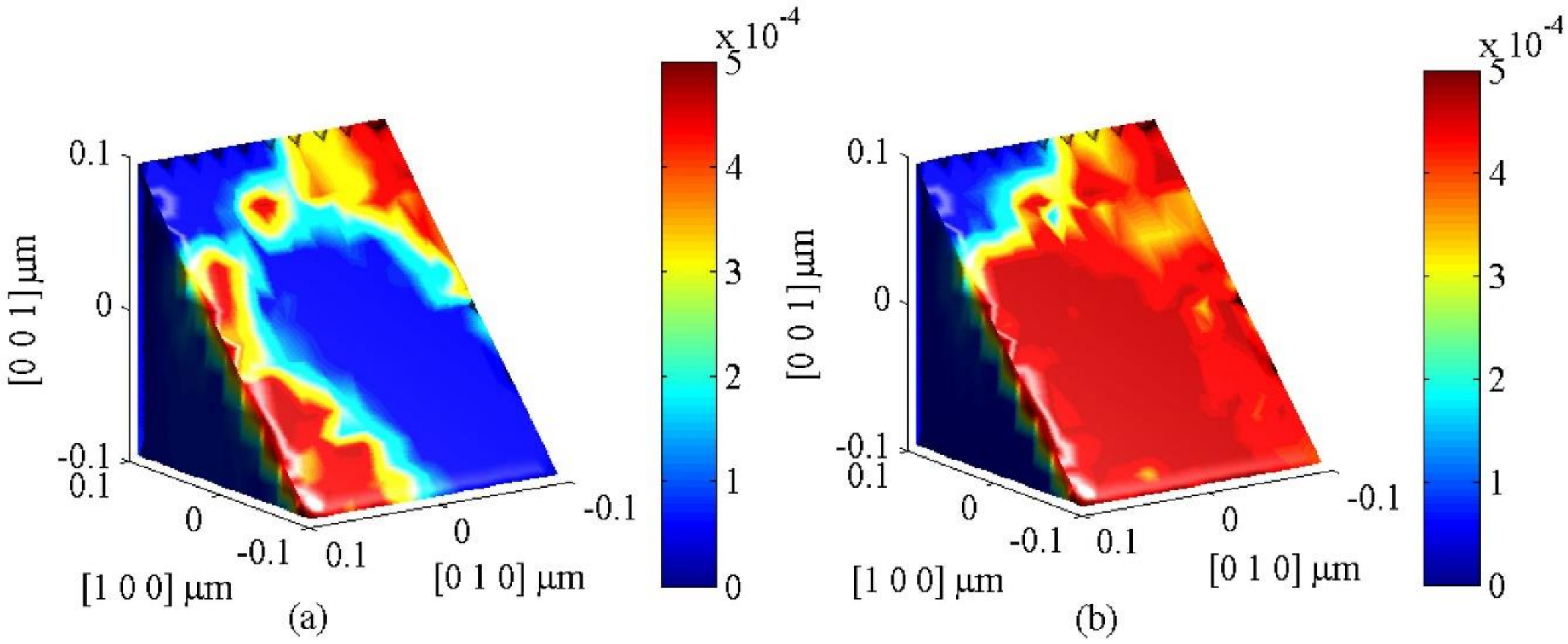

Fig. 9. Local effective plastic strain in the dislocation glide plane, (a) impenetrable precipitate (Fig. 8a), (b) penetrable precipitate (Fig. 8b).

## 4. Conclusion

We computationally studied the short-range interactions of an edge dislocation with an array of equally-spaced identical precipitates. Although the dislocation line is initially an edge type, during the bypass, it turns into a mixed edge and screw dislocation. We incorporated the precipitate resistance into a dislocation dynamics approach as a frictional force against dislocation movement. We quantified the dislocation curvature and angle change during the bypass phenomenon for various levels of applied shear stress and precipitate resistance. The results show that the maximum curvature of a dislocation during the bypass is directly related to precipitate resistance. A dislocation line undergoes a higher level of deformation when it encounters a stronger precipitate. The study of local effective plastic strains shows that an increase in precipitate resistance causes a significant decrease in the effective plastic strain. However, if a dislocation passes through a shearable precipitate, effective plastic strain is independent of precipitate resistance and the applied stress. A multi-scale analysis of plastic deformation close to precipitates shows a high strain gradient at the interface of non-shearable precipitates.